\documentclass[aps,twocolumn,nofootinbib]{revtex4}
\usepackage{graphicx}
\usepackage{amsfonts}
\usepackage{amssymb}
\usepackage{amsbsy}
\usepackage{amsmath}
\usepackage{mathrsfs} 
\usepackage{latexsym}
\usepackage{natbib}
\usepackage{bm}
\usepackage{subfigure}
\usepackage{color}

\def\ce{\mathrm{ce}}
\def\se{\mathrm{se}}

\def\k{\mathbf{k}}

\def\x{\mathbf{x}}
\def\y{\mathbf{y}}
\def\xiv{\vec{\xi}}

\def\k{\mathrm{k}}

\def\H{\mathcal{H}}

\def\lstar{l_{\star}}

\begin{document}
\title{Response of a uniformly accelerated Unruh-DeWitt detector 
 in polymer quantization}

\author{Gopal Sardar}
\email{gopalsardar.87@gmail.com}
 
\author{Subhashish Banerjee}
\email{subhashish@iitj.ac.in}

\affiliation{ Indian Institute of Technology Jodhpur, Jodhpur 342011, India}

\date{\today}

\begin{abstract}
If an Unruh-DeWitt detector moves with a uniform acceleration in Fock-space 
vacuum, then the transition rate of the detector is proportional to the thermal 
spectrum. It is well known that the transition rate of the detector crucially 
depends on the two-point function along the detectors trajectory and in order to 
compute it the standard ``$i \epsilon$'' regularization is used for Fock space. 
Numerically, we show here that the regulator $\epsilon$ is generic in polymer 
quantization, the quantization method used in \emph{loop quantum gravity} with a 
finite value $\epsilon \approx 2.16$, which leads to non-thermal spectrum for 
the uniformly accelerated detector. We also discuss the response of a spatially 
smeared detector.

\end{abstract}

\maketitle

\section{Introduction}

Minkowski vacuum is perceived by an observer accelerating with a uniform 
acceleration $a$ as a thermal distribution with temperature $T = a/2\pi k_B$, 
the Unruh temperature, where $k_B$ is the \emph{Boltzmann constant} 
\cite{Fulling:1972md,Unruh:1976db,Crispino:2007eb,de2006unruh,Takagi01031986, 
Longhi:2011zj,Davies:1974th,Birrell1984quantum,Banerjee:2015bxl,Omkar:2014hba,
Banerjee:2016ahn}.
The Unruh effect can be approached from different perspectives. A common 
approach is the application of  \emph{Bogoliubov transformations}. The 
expectation value of the number density operator in the Fock vacuum state, as 
perceived by an accelerated observer, has the form of a blackbody distribution 
at Unruh temperature. The Unruh effect appears to 
explicitly depend on the contributions from trans-Planckian modes, as observed 
by an \emph{inertial} observer. This thus provides a potential candidate for 
exploring the implications of possible Planck-scale physics 
\cite{Nicolini:2009dr,Padmanabhan:2009vy,Agullo:2008qb,Chiou:2016exd,
Alkofer:2016utc}, such as those falling within the ambit of quantum gravity.

Another approach, adopted here, is to compute the response function of the  
Unruh-DeWitt 
detector moving along the trajectory of an accelerated observer 
\cite{DeWitt:1980hx,Hinton_1983dq,Schlicht:2003iy,Louko:2006zv, 
Unruh:1983ms,Louko:2014aba,Sriramkumar:1994pb,Agullo:2010iq,Fewster:2015dqb,
Satz:2006kb,Langlois:2005nf,Hummer:2015xaa}.
In this approach, one considers a two-level quantum mechanical detector which 
weakly couples to the ambient environment, such as a scalar matter field. By 
computing the transition 
probability of the detector between the energy levels and comparing with the 
\emph{spontaneous} and \emph{induced} 
emission or absorption, one can understand the state of 
the scalar matter field. In particular, the detector response function depends 
upon the Wightman (two-point) function of the scalar field.

Polymer (loop) quantization \cite{Ashtekar:2002sn,Halvorson-2004-35} is used as 
a quantization technique in loop quantum gravity 
\cite{Ashtekar:2004eh,Rovelli2004quantum,Thiemann2007modern}. It has an inbuilt 
(dimension-full) parameter apart from the Planck constant $\hbar$. This new 
scale corresponds to Planck length $L_p$ in the context of full quantum gravity. 
Further, here both position and momentum operators cannot be simultaneously 
defined. These features make polymer quantization unitarily inequivalent to 
Schr\"odinger quantization \cite{Ashtekar:2002sn}. Here  we compute the response 
function of an Unruh-DeWitt detector in the context of polymer quantization of 
scalar field weakly coupled to the detector.

The plan of the paper is as follows.
In section \ref{Sec:RindlerObserver}, we briefly discuss about spacetime 
as seen by a  uniformly accelerating observer in Minkowski spacetime $i.e.,$ the 
Rindler observer and its trajectory. Next,
in section \ref{Sec:MasslessScalarField}, we consider polymer quantization of a 
massless free scalar 
field in the canonical approach. The properties of the Unruh-DeWitt 
detector are then studied. Subsequently, we study the behaviour of the Fock 
space 
two-point 
function analytically by considering the standard ``$i \epsilon$'' 
regularization. By comparing the numerically computed polymer and Fock space 
two-point functions we show that the regulator $\epsilon$, 
used for the standard regularization for Fock space, is 
generic in the case of polymer-two-point function with a finite value $\epsilon 
\approx 2.16$. Thus, a \emph{generic cut-off} is seen to emerge in polymer 
quantization. Then we compute the induced transition rate of the 
Unruh-DeWitt detector along the Rindler trajectory. We show that, in Fock 
quantization, the induced transition rate is proportional to Planck 
distribution. However, in polymer quantization, due to the large value of the 
generic regulator $\epsilon$, the induced transition rate deviates from the 
Planck distribution. Next, we compute the induced transition rate by 
considering spatially smeared detector in both Fock and polymer 
quantizations. Finally, we make our conclusions.


\section{Rindler Spacetime}\label{Sec:RindlerObserver}

The spacetime of an observer who is moving with a \emph{uniform acceleration} in 
Minkowski spacetime can be described by the so-called Rindler metric. Using 
\emph{conformal} Rindler coordinates $\bar{x}^{\alpha} = (\tau,\xi,y,z) \equiv 
(\tau,\xiv)$ together with natural units ($c=\hbar=1$) the Rindler metric can be 
expressed as \cite{Rindler:1966zz}
\begin{equation}
 \label{RindlerMetric}
  ds^2 = e^{2a\xi} \left( -d\tau^2 + d\xi^2 \right) + dy^2 + dz^2
  \equiv g_{\alpha\beta}d\bar{x}^{\alpha} d\bar{x}^{\beta}  ~,
\end{equation}
 where the parameter $a$ is the magnitude of \emph{acceleration} 4-vector. With 
respect to an \emph{inertial} observer the Minkowski metric with Cartesian 
coordinates $x^{\mu} = (t,x,y,z) \equiv (t,\x)$ would appear as $ds^2 = 
\eta_{\mu\nu}dx^{\mu}dx^{\nu} = - dt^2 + dx^2 + dy^2 + dz^2$.
If the uniformly accelerated 
observer \emph{i.e.,} Rindler observer moves along positive x-axis with 
respect to the inertial observer, the coordinates are related each other by 
\begin{equation}
 \label{RindlerMinkowskiRelation}
 t =  \frac{1}{a} e^{a\xi} \sinh a\tau ~,~~~
 x =  \frac{1}{a} e^{a\xi} \cosh a\tau ~,
\end{equation}
 Here, $y$ and $z$ coordinates are the unaffected. 
 It can be seen from the equation (\ref{RindlerMinkowskiRelation}) 
that only a wedge-shaped section of Minkowski spacetime is covered by Rindler 
spacetime, called the \emph{Rindler wedge}.

\section{Scalar field}
\label{Sec:MasslessScalarField}

We 
consider a massless scalar field $\Phi(x)$ weakly coupled to the 
detector in Minkowski spacetime. The action for corresponding scalar field 
dynamics is  
\begin{equation}
 \label{ScalarActionMinkowski}
 S_{\Phi} = \int d^4x \left[ - \frac{1}{2} \sqrt{-\eta} \eta^{\mu\nu}  
 \partial_{\mu} \Phi(x)  \partial_{\nu} \Phi(x) \right] ~.
\end{equation}
 The field Hamiltonian is 
\begin{equation}\label{SFHamGen}
H_{\Phi}  =  \int d^3\x \left[ \frac{\Pi^2}{2\sqrt{q}} +
\frac{\sqrt{q}}{2} q^{ab} \partial_a\Phi \partial_b\Phi
\right] ~,
\end{equation}
where $q_{ab}$ is the metric on \emph{spatial} hyper-surfaces labeled by $t$. 
Poisson bracket between the field $\Phi = \Phi(t,\x)$ and its conjugate 
field momentum $\Pi = \Pi(t,\x)$ is 
\begin{equation}\label{PositionSpacePB}
\{\Phi(t,\x), \Pi(t,\y)\} = \delta^{3}(\x-\y) ~,
\end{equation}
where $\delta^{3}(\x-\y)$ is the Dirac delta.

\subsection{Fourier modes}

 Here we express Fourier modes for the scalar field and its 
conjugate field momentum as
\begin{equation}\label{FourierModesDef}
\Phi = \frac{1}{\sqrt{V}} \sum_{\k} \tilde{\phi}_{\k}(t) e^{i \k\cdot\x} ,~
\Pi  = \frac{1}{\sqrt{V}} \sum_{\k} \sqrt{q} ~\tilde{\pi}_{\k}(t) 
e^{i \k\cdot\x},
\end{equation}
where $V=\int d^3\x \sqrt{q}$ is the spatial volume. In Minkowski spacetime, as 
the space is non-compact, the spatial volume would normally diverge. In order 
to avoid this issue, it is convenient to use a fiducial box of finite 
volume. Kronecker and Dirac delta functions then can be expressed as $\int 
d^3\x 
\sqrt{q} ~e^{i (\k-\k')\cdot \x} = V \delta_{\k,\k'}$ and $\sum_{\k} e^{i 
\k\cdot (\x-\y)} = V \delta^3 (\x-\y)/\sqrt{q}$, respectively. 
The field Hamiltonian (\ref{SFHamGen}) can be expressed as $H_{\Phi} = 
\sum_{\k} \H_{\k}$, where Hamiltonian density for the $\k-$th Fourier mode is
\begin{equation}\label{SFHamFourierMinkowski}
\H_{\k} = \frac{1}{2} \tilde{\pi}_{-\k} \tilde{\pi}_{\k} +
\frac{1}{2} |\k|^2 \tilde{\phi}_{-\k}\tilde{\phi}_{\k}  ~.
\end{equation}
Poisson brackets between these Fourier modes and their conjugate momenta 
are given by
\begin{equation}\label{Minkowski:MomentumSpacePB}
\{\tilde{\phi}_{\k}, \tilde{\pi}_{-\k'}\} =  \delta_{\k,\k'}
~.
\end{equation}

One usually redefines the \emph{complex-valued} modes $\tilde{\phi}_{\k}$ and 
momenta $\tilde{\pi}_{\k}$ in terms of the real-valued functions $\phi_{\k}$ and 
$\pi_{\k}$ in order to satisfy the \emph{reality condition} of the scalar field 
$\Phi$. Therefore, the corresponding Hamiltonian density and Poisson brackets 
become
\begin{equation}\label{SFHamFourierMinkowskiReal}
\H_{\k} = \frac{1}{2} \pi_{\k}^2 + \frac{1}{2} |\k|^2 \phi_{\k}^2
~~~;~~~ \{\phi_{\k},\pi_{\k'}\} =  \delta_{\k,\k'} ~,
\end{equation}
which is the standard Hamiltonian for a system of decoupled harmonic 
oscillators. The relevant eigenvalue equation is 
$\hat{\H}_{\k}|n_{\k}\rangle = E_n^{(\k)}|n_{\k}\rangle$, where $|0_\k\rangle$ 
is the vacuum state 
of the $k^{th}$ mode. The vacuum state of the scalar field is
 $|0\rangle=\Pi_{\k}\otimes |0_\k\rangle$.

\section{Unruh-DeWitt Detector}
\label{Sec:Unruh-DeWitt-Detector}

Unruh-Dewitt detector is a point-like quantum mechanical 
system which has two internal energy levels. Here we study the response 
function of the Unruh-DeWitt detector which interacts 
\emph{weakly} with the scalar field via a \emph{linear} coupling. The 
energy eigenvalue equation of the detector is 
\begin{equation}\label{DetectorEigenvalueEquation}
\hat{H}_0 |g\rangle = \omega_g|g\rangle ~~;~~
\hat{H}_0 |e\rangle = \omega_e|e\rangle ~,
\end{equation}
where $\hat{H}_0$ is the unperturbed Hamiltonian of the detector and 
$|g\rangle$ and $|e\rangle$ represent the ground and 
excited states, respectively. The energy gap is 
\begin{equation}\label{DetectorEnergyGap}
\omega \equiv \left(\omega_e - \omega_g \right) > 0 ~.
\end{equation}
The interaction Hamiltonian of the Unruh-DeWitt detector is taken as
\begin{equation}\label{DetectorFieldInteractionHamiltonian}
\hat{H}_{int} (\tau) = \lambda ~\hat{\mu}(\tau) \hat{\Phi}(x(\tau)) ~,
\end{equation}
where $\lambda$ denotes the \emph{coupling constant} and $\hat{\mu}(\tau)$ is 
the monopole moment operator of the detector. The detector's trajectory  
$x^{\mu}(\tau)$ is parametrized using the proper time $\tau$. Hence, the 
total detector Hamiltonian is 
\begin{equation}\label{TotalSystemHamiltonian}
\hat{H} = \hat{H}_0 + \hat{H}_{int}  ~.
\end{equation}
It is convenient to work in the 
interaction picture, in which the time evolution operator of the 
detector is given by
\begin{equation}\label{EvolutionOperator}
U(\tau_f,\tau_i) =  1 - i \int_{\tau_i}^{\tau_f} d\tau' U(\tau',\tau_i)
\hat{H}_I (\tau') ~.
\end{equation}
If the state of the scalar field is $|\Xi_{\tau}\rangle$, the combined state of 
the detector and the scalar 
field at a given proper time $\tau$ is
\begin{equation}\label{GeneralCombinedState}
|\psi, \Xi; \tau\rangle \equiv |\psi_{\tau}\rangle_I \otimes
|\Xi_{\tau}\rangle  ~.
\end{equation}
The transition \emph{amplitude} from the state $|g, \Xi_i; 
0\rangle$ to  $|e, \Xi_f; \tau\rangle$ can be expressed as
\begin{equation}\label{TransitionAmplitude}
\mathop{Amp} = - i \lambda \int_{0}^{\tau} d\tau' 
\langle e, \Xi_f; \tau| \hat{\mu}_I(\tau') \hat{\Phi}(x(\tau')) 
|g, \Xi_i;0\rangle ~,
\end{equation}
where $\hat{\mu}_I(\tau)$ is the monopole moment operator in the interaction 
picture and the corresponding \emph{probability} of transition is
\begin{equation}\label{TransitionProbability}
\mathcal{P}_{|g, \Xi_i; 0\rangle \to |e, \Xi_f; \tau\rangle} = |Amp |^2 ~.
\end{equation}
Now if  the detector initially is at ground state and the scalar field is in its 
vacuum state \emph{i.e.,} $|\Xi_i\rangle = |0\rangle$, then the transition 
probability at a time $\tau$ for the detector being in the excited state 
$|e\rangle$ for all possible field states is
\begin{equation}\label{TransitionProbabilityDetector}
\mathcal{P}_{\omega}(\tau,0) \equiv \mathcal{P}_{|g; 0\rangle \to |e; 
\tau\rangle} = 
\sum_{\{|\Xi_f\rangle\}} \mathcal{P}_{|g, \Xi_i; 0\rangle \to |e, \Xi_f; 
\tau\rangle} ~.
\end{equation}
The transition probability (\ref{TransitionProbabilityDetector}) can be 
easily expressed in the form of
\begin{equation}\label{TransitionProbabilityDetector2}
\mathcal{P}_{\omega}(\tau,0) = A_0 \mathcal{F}_{\omega}(\tau,0) ~,
\end{equation}
where $A_0 = \lambda^2 |\langle e| \hat{\mu}(0) |g\rangle|^2$. $A_0$ depends on 
the internal properties of the detector system.  
$\mathcal{F}_{\omega}(\tau,0)$ 
is the \emph{response function} of the detector, and is 
\begin{equation}\label{DetectorResponseFunction}
\mathcal{F}_{\omega}(\tau,0) = \int_{0}^{\tau} \int_{0}^{\tau} d\tau' d\tau''
e^{-i\omega(\tau''-\tau')} ~G(\tau'',\tau') ~,
\end{equation}
where $G(\tau'',\tau')$ is the \emph{two-point function} of the scalar field
which can be expressed as
\begin{equation}\label{TwoPointFunction}
G(\tau'',\tau') = G(\tau''-\tau')  =
\langle 0| \hat{\Phi}(x(\tau'')) \hat{\Phi}(x(\tau')) |0\rangle ~.
\end{equation}

 We can now define the
\emph{instantaneous} transition rate  by 
using equation (\ref{TransitionProbabilityDetector2}) with a scaling,
as
\begin{equation}\label{DetectorTransitionRate}
R_{\omega}(\tau,0) \equiv \left(\frac{2\pi}{A_0} \right)
\frac{d \mathcal{P}_{\omega}}{d\tau}
= 2\pi \int_{-\tau}^{\tau} d\tau' e^{-i\omega \tau'} G(\tau') ~,
\end{equation}
which can be re-expressed as
\begin{equation}\label{DetectorTransitionRate2}
R_{\omega}(\tau,0) = R_{\omega}^0 + \Delta R_{\omega}(\tau) ~.
\end{equation}
The $R_{\omega}^0$ denotes the time independent part of the induced transition 
rate \emph{i.e.}, the \emph{non-transient} part, which can be expressed as
\begin{equation} \label{IntegralR0}
R_{\omega}^0 =  2\pi \int_{-\infty}^{\infty} d\tau' 
e^{-i\omega \tau'} G(\tau') ~.
\end{equation}
On the other hand, $\Delta R_{\omega}(\tau)$ denotes time-dependent 
\emph{i.e.}, \emph{transient} part of the induced transition rate and it 
can be expressed as 
\begin{equation} \label{IntegralDeltaR}
\Delta R_{\omega}(\tau) = - 2\pi \int_{\tau}^{\infty} d\tau' 
\left[ e^{-i\omega \tau'} G(\tau') +  e^{i\omega \tau'} G(-\tau') \right] ~.
\end{equation}
 $\Delta R_{\omega}(\tau) \to 0$  with increasing observation time   
\emph{i.e.}, $\tau \to \infty$.

\subsection{Two-point function}\label{sec:two-point-function}

We can see from the equation (\ref{DetectorTransitionRate}) that induced 
transition rate of the Unruh-DeWitt detector is completely determined from the 
properties of the two-point function. The general form of two-point 
function can be written in terms of the Fourier modes (\ref{FourierModesDef}) as
\begin{equation}
\label{MinkowskiTwoPointDef2}
G(x,x') = \langle 0|\hat{\Phi}(x) \hat{\Phi}(x')|0\rangle =
\frac{1}{V} \sum_{\k} D_{\k}(t,t') e^{i {\k}\cdot(\x-\x')} ,
\end{equation}
where the matrix element $D_{\k}(t,t')$ is given by
\begin{equation}\label{DkDefinition}
D_{\k}(t,t') = 
\langle 0_{\k}| e^{i\hat{\H}_{\k}t} \hat{\phi}_{\k} e^{-i\hat{\H}_{\k}t}
e^{i\hat{\H}_{\k}t'} \hat{\phi}_{\k} e^{-i\hat{\H}_{\k}t'}
|0_{\k}\rangle.
\end{equation}
Exploiting the independence of  
 the Hamiltonians and the corresponding Poisson 
brackets (\ref{SFHamFourierMinkowskiReal}) from the fiducial 
volume, the discrete summation over the modes can be replaced by an 
integration. Thus, the two-point function (\ref{MinkowskiTwoPointDef2}) can be 
expressed as
\begin{equation}
\label{MinkowskiTwoPointIntegralDef}
G(x,x') = \int \frac{d^3\k}{(2\pi)^3} ~  D_{\k}(t,t') 
~e^{i {\k}\cdot(\x-\x')} ~.
\end{equation}
By expanding the state $\hat{\phi}_{\k} |0_{\k}\rangle$ in the basis of energy 
eigenstates as $\hat{\phi}_{\k}|0_{\k}\rangle = \sum_{n} b_n |n_{\k}\rangle$ 
and 
using energy spectrum of the Fourier modes, the matrix element $D_{\k}(t,t')$ 
can be expressed as
\begin{equation}
\label{DkFunctionGeneral}
D_{\k}(t-t') \equiv D_{\k}(t,t') = \sum_{n} |b_n|^2 e^{-i\Delta E_n (t-t')},
\end{equation}
where $\Delta E_n \equiv E_n^{(\k)} - E_0^{(\k)}$ and
$b_n = \langle n_{\k}| \hat{\phi}_{\k} |0_{\k}\rangle$. We can see that the 
matrix 
element $D_{\k}(t-t')$ depends only on magnitude $|\k|$. Therefore, one can 
carry out the angular integration using \emph{polar coordinates}, and the 
reduced two-point function (\ref{MinkowskiTwoPointIntegralDef}) can be 
expressed as
\begin{equation}
\label{KGPropagatorDiffPM}
G(x,x') = G_{+} - G_{-} ~,
\end{equation}
where
\begin{equation}
\label{GPMDefinition}
G_{\pm} =  \frac{i}{4\pi^2|\Delta\x|} \int dk ~k D_{k}(\Delta t) 
~e^{\mp i k |\Delta \x|} ~,
\end{equation}
with $k = |\k|$, $\Delta \x = \x-\x'$ and $\Delta t = t-t'$.

\section{Fock quantization: Detector Response}

In Fock quantization, the energy spectrum and the 
coefficients $b_n$ (\ref{DkFunctionGeneral}) of the Fourier modes 
(\ref{SFHamFourierMinkowskiReal}) of the scalar field
 are given by
\begin{equation}\label{FockSpectrum}
E^n_{\k} = \left(n+\frac{1}{2}\right)|\k| ~~;~~ 
\Delta E_n = n |\k|   ~~;~~
 b_n = \frac{\delta_{1,n}}{\sqrt{2|\k|}} ~.
\end{equation}
Using above equation (\ref{FockSpectrum}), the two-point function 
(\ref{KGPropagatorDiffPM}) then reduces to its standard 
form 
\begin{equation}
\label{FockTwoPointFunction}
G(x,x') = \frac{1}{4\pi^2 \left[-(\Delta t - i\epsilon)^2 
+ |\Delta \x|^2\right]}~,
\end{equation}
where $\Delta x^2 = - \Delta t^2 + |\Delta \x|^2$ is the Lorentz invariant 
spacetime interval. Further, $\epsilon$ is a small, positive parameter that is 
introduced as the standard integral regulator, $i.e$, as an additional term 
$e^{-\epsilon k }$ in the integrand of Eq. (\ref{GPMDefinition}).

Now we compute the transition rate of an Unruh-DeWitt detector which 
moves along a Rindler trajectory given by $x_d^\mu (\tau) = 
(\sinh(a\tau)/a,\cosh(a\tau)/a,0,0)$. Defining a new variable $\rho = e^{a 
\tau}$,  the time interval $\Delta t$ and spatial separation $|\Delta \x|$ can 
be expressed as
\begin{equation}\label{RindlerDeltaXandT}
\Delta t = \frac{(\rho^2-1)}{2 a\rho} ~~,~~
|\Delta \x| = \frac{(\rho-1)^2}{2 a\rho} ~,
\end{equation}
and the corresponding two-point function (\ref{FockTwoPointFunction}) becomes
\begin{equation}\label{FockTwoPointFunctionRindlerEta}
G(\rho) =- \frac{a^2~ \rho}
{4 \pi^2 \left(\rho - 1 -i\epsilon \right)^2}  ~.
\end{equation}
Then the non-transient part of the transition rate (\ref{IntegralR0}) can be 
expressed as
\begin{equation}\label{IntegralR0FockRindler}
R_{\omega}^0 = -\frac{a}{2\pi} \int_{0}^{\infty} d\rho 
\frac{\rho^{-i\omega/a}} {\left(\rho - 1 - i\epsilon\right)^2} 
\equiv \int_{0}^{\infty} d\rho~ h(\rho) ~.
\end{equation}
We can see that the integrand $h(\rho)$ has a \emph{pole of second order} at 
$\rho=1+i\epsilon$ and it is also a \emph{multi-valued} function of complex 
variable $\rho$. 
\begin{figure}[h!]
 \includegraphics[scale=.5]{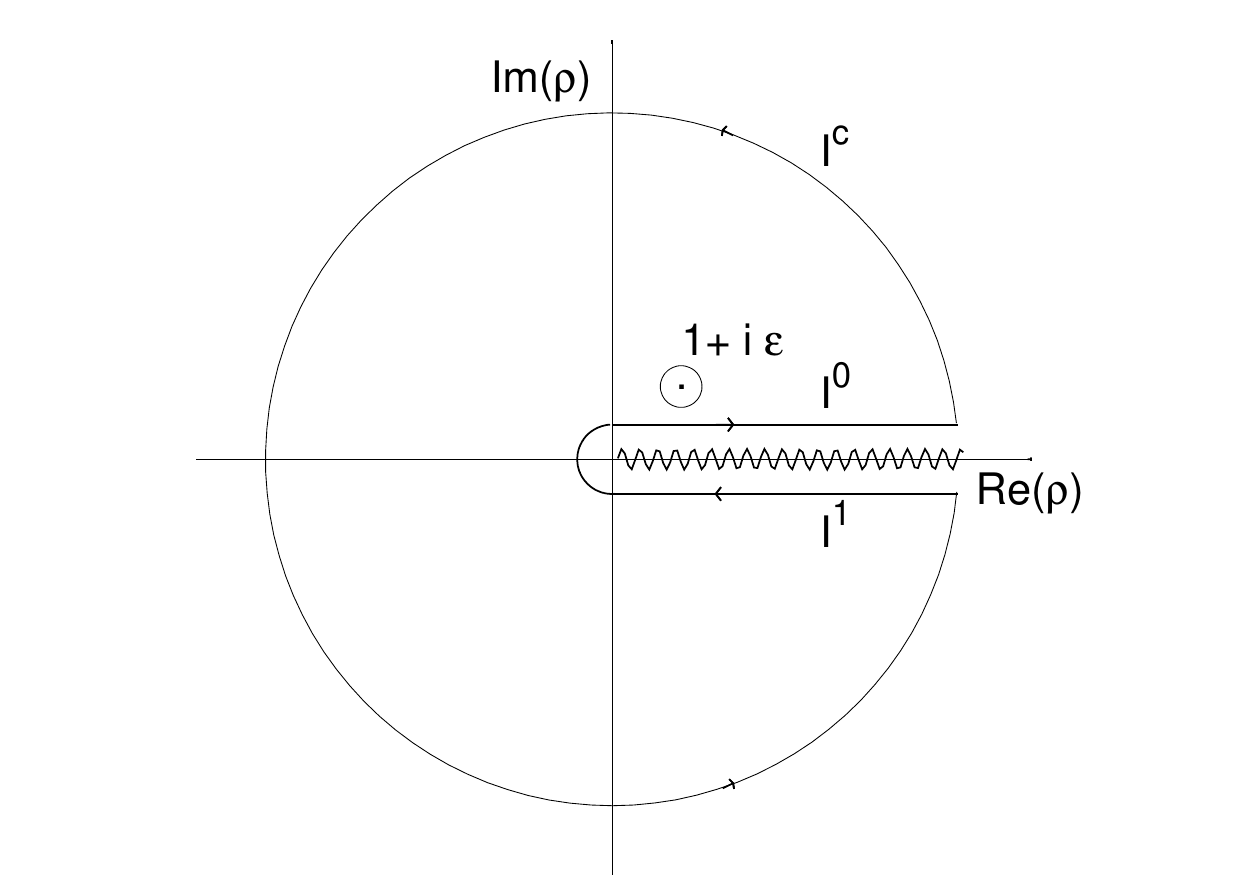}
 \caption{\label{fig:contour}Contour used to evaluate $R_{\omega}^0$}
\end{figure}
Following the contour, as shown in Fig. \ref{fig:contour}, 
the contour integral can be expressed as
\begin{equation}\label{IntegralI0FockRindler}
\oint d\rho~ h(\rho) = 
I^0 + I^1 + I^c = (2\pi i)~
\mathrm{Res}[h(\rho)]_{|\rho=1+i\epsilon} ~,
\end{equation}
where $\mathrm{Res}[h(\rho)]_{|\rho=1+i\epsilon}$ is the \emph{residue} of the 
function $h(\rho)$ evaluated at the pole $\rho=(1+i\epsilon)$. It can be 
easily 
shown that $I^c =0$ and $I^1 = - e^{2\pi\omega/a}~ I^0$. Therefore, the 
non-transient part of the transition rate can be expressed as
\begin{equation}\label{RindlerGeneralR0}
R_{\omega}^0  = I^0 =  
\frac{-(2\pi i)\mathrm{Res}[h(\rho)]_{|\rho=1+i\epsilon}}{e^{2\pi\omega/a} - 1} 
~.
\end{equation}
After taking the limit $\epsilon \to 0$, the evaluated residue at the pole of 
the two-point function in Fock space will be
\begin{equation}\label{FockRindlerResidue}
\mathrm{Res}[h(\rho)]_{|\rho=1+i\epsilon} = \frac{i \omega}{2\pi}  ~,
\end{equation}
which leads the induced transition rate to become
\begin{equation}\label{RindlerFockR0Final}
R_{\omega}^0  = \frac{\omega}{e^{2\pi\omega/a} - 1} ~.
\end{equation}
This is the standard expression for mean energy per mode of a system in thermal 
equilibrium at the Unruh temperature $T = a/2\pi k_B$.

By considering reasonably large but finite time of observation, the transient 
part $\Delta R_{\omega}$ will be
\begin{equation}\label{IntegralDeltaIFockRindler}
\Delta R_{\omega}({\tau}) \approx \frac{e^{-a\tau}}
{\pi \left( 1+ \omega^2/a^2 \right)} 
\left[a \cos(\omega \tau) -\omega \sin(\omega\tau)\right].
\end{equation}
It clearly shows that the transient terms decay 
exponentially. 

Therefore, if an Unruh-DeWitt detector is on for a sufficiently long time along 
a Rindler trajectory having magnitude of the acceleration 4-vector $a$, then 
Minkowski vacuum will appear as a thermal state at the temperature $T = a/2\pi 
k_B$.

\section{Polymer quantization}
In this section we discuss briefly about polymer quantization and then study 
Unruh effect numerically. In polymer quantization, the position 
operator $\hat{x}$ and translation operator $\hat{U}(\lambda)$ are considered 
as 
basic operators. In Polymer Hilbert space, the momentum operator does not exist 
as the translation operator is not weakly continuous in the parameter 
$\lambda$. However, one can define an analogous momentum operator as $ 
\hat{p}_{\lambda}=1/(2i\lambda)(\hat{U}(\lambda)-\hat{U}(-\lambda))$, such that 
 the usual momentum operator is recovered in the limit $\lambda 
\to 0$. In polymer quantization this limit does not exist and 
$\lambda$ is considered as a small and finite scale, 
$\lambda_{\star}$.  This dimension-full parameter is analogous to Planck-length 
$L_p$ as $\lambda_{\star} \sim \sqrt{L_p}$.

The energy spectrum of the $\k-$th oscillator is
\cite{Hossain:2010eb}  
\begin{equation}
 \label{EigenValueMCFRelation}
 \frac{E_{\k}^{2n}}{|\k|} = \frac{1}{4g} + \frac{g}{2} ~A_n(g)  ~,~
 \frac{E_{\k}^{2n+1}}{|\k|} = \frac{1}{4g} + \frac{g}{2} B_{n+1}(g) ~,
\end{equation}
where $n\ge0$, $A_n$, $B_n$ are Mathieu characteristic value
functions and $g =|\k|{\lambda_{\star}}^2\equiv |\k|~\lstar$ is a 
dimension-less
parameter. In terms of the cosine and sine elliptic functions 
\cite{Abramowitz1964handbook} $\ce_n$ and $\se_n$, respectively,
the energy eigenstates can be expressed as
\begin{eqnarray}
 \psi_{2n}(v) &=& \ce_n(1/4g^2,v)/\sqrt{\pi}, \\ \nonumber
\psi_{2n+1}(v) &=& \se_{n+1}(1/4g^2,v)/\sqrt{\pi},
\end{eqnarray}
where 
$v = \pi_{\k} \sqrt{\lstar} + \pi/2 $.
 Superselection 
rules are invoked to arrive at these $\pi$-periodic and $\pi$-antiperiodic 
states in $v$ \cite{Barbero:2013lia}.

For low-energy modes \emph{i.e.}, for small $g$, the energy spectrum 
(\ref{EigenValueMCFRelation}) can be expressed as the energy spectrum of a
regular harmonic oscillator along with perturbative corrections as
\begin{equation}
 \label{EEvalueSmallg}
  \frac{E_{\k}^{2n}}{|\k|} \approx \frac{E_{\k}^{2n+1}}{|\k|} \approx
   \left(n+\frac{1}{2}\right) + \mathcal{O}(g)~.
\end{equation}
Therefore, one can recover the standard energy spectrum of the harmonic 
oscillator in the limit $g \to 0$. However, we should emphasize here that 
polymer energy spectrum has two-fold degeneracy as $g \to 0$ 
and it is lifted for finite values of $g$. The coefficients  $b_{4n+3} = i 
\sqrt{\lstar} \int_0^{2\pi} \psi_{4n+3} \partial_v\psi_{0} dv$ are non-vanishing 
in polymer quantization for all $n=0, 1, 2, \ldots$, whereas in Fock 
quantization, only one $b_n$ is non-vanishing (\ref{FockSpectrum}). 
Using asymptotic expressions of Mathieu functions, one can approximate
the energy gaps $\Delta E_n$ between the levels and coefficients $b_{4n+3}$ for 
low-energy modes (sub-Planckian), $g\ll 1$, 
\begin{equation}\label{SmallgDeltaE4n+3}
\frac{\Delta E_{4n+3}}{|\k|} = (2n+1) - \frac{(4n+3)^2-1}{16} g  + 
\mathcal{O} \left( g^2 \right) ~,
\end{equation}
for $n \ge 0$, and 
\begin{equation}
\label{SmallgC4n+3}
  b_3 = \frac{i}{\sqrt{2|\k|}} \left[1 
  +\mathcal{O}\left(g\right) \right] ~,~
  \frac{b_{4n+3}}{b_{3}} = \mathcal{O}\left(g^n\right),
\end{equation}
for $n >0$. On the other hand, for high energy modes (super-Planckian), 
 $g\gg1$, 
\begin{equation}
 \label{LargegDeltaE4n+3}
 \frac{\Delta E_{4n+3} }{|\k|}  =  2(n+1)^2 g + 
  \mathcal{O}\left(\frac{1}{g^3}\right) ,
\end{equation}
for $n \ge 0$, and 
\begin{equation}
  \label{LargegC4n+3}
  b_{3} =  i\sqrt{\frac{g}{2|\k|}} \left[\frac{1}{4g^2} +
  \mathcal{O}\left(\frac{1}{g^6}\right) \right],
  \frac{b_{4n+3}}{b_{3}} = \mathcal{O}\left(\frac{1}{g^{2n}}\right),
\end{equation}
for $n > 0$. Therefore, in polymer quantization, we can approximate matrix 
element $D_{\k}(\Delta t)$ (\ref{DkFunctionGeneral}) by taking only the first 
element as
\begin{equation}
\label{DkPolyApprox1}
D_{\k}^{poly}(\Delta t) \simeq  |b_3|^2 e^{ -i \Delta E_3 \Delta t} ~,
\end{equation}
for both the cases.

\subsection{Numerical computation of two-point function and Unruh effect}
Using asymptotic expressions of the $b_3$ and $\Delta E_3$ one could 
analytically evaluate the two-point function for asymptotic spacetime 
intervals in polymer quantization. However, for all possible spacetime intervals
it does not appear to be possible in polymer quantization, as in Fock 
quantization. In order to obtain a comprehensive picture of two-point 
function for all possible spacetime intervals we use numerical 
techniques.

\subsubsection{Matrix element $D_{\k}(\Delta t)$}
\begin{figure}[h!]
\centering
 \includegraphics[scale=0.7]{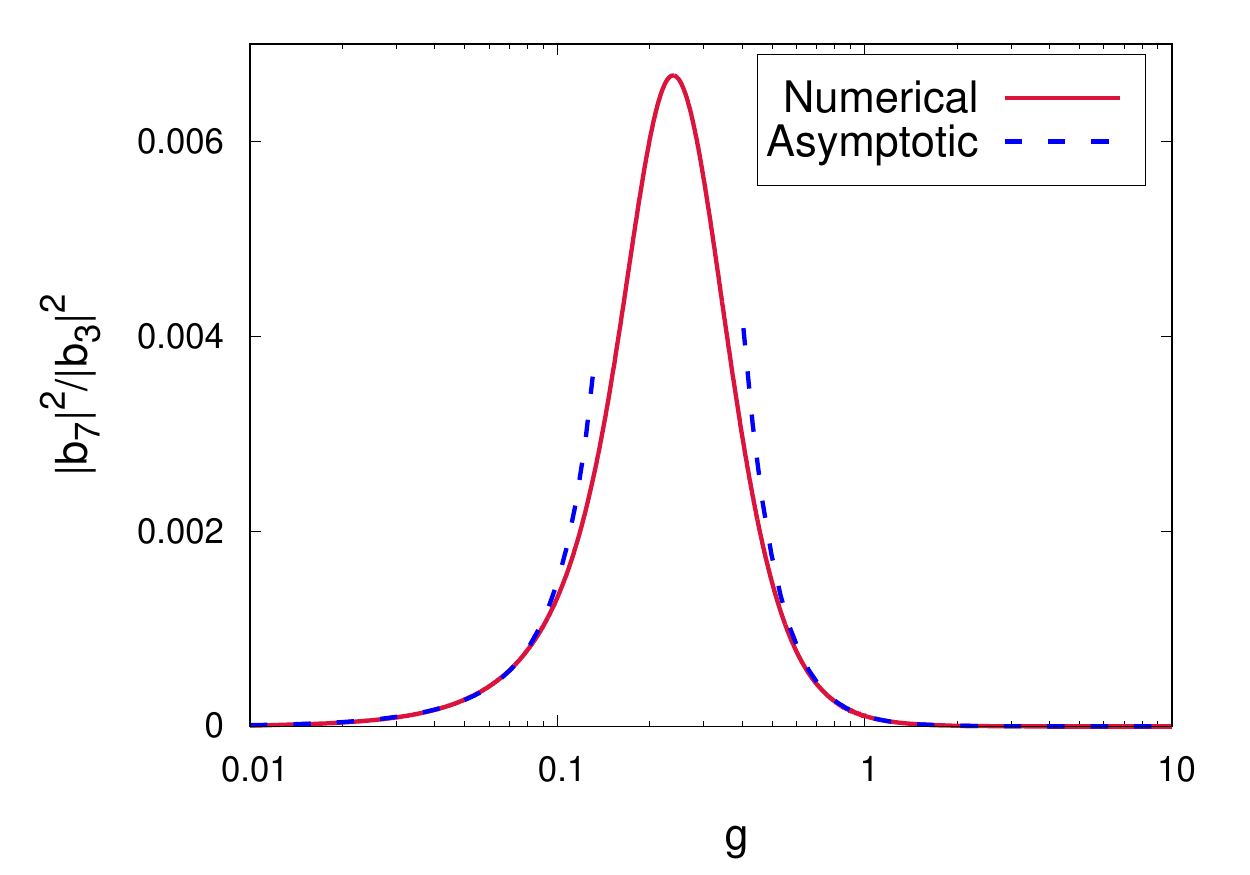}
 \caption{The solid red line represents numerically evaluated ratio between 
$|b_7|^2$ and $|b_3|^2$. The blue dashed line represents the same ratio using
their respective asymptotic expressions.  
}
 \label{fig:c3c7comparison}
\end{figure}

 The two-point 
function is completely understood from the matrix element $D_k(\Delta t)$, cf. 
 equation (\ref{GPMDefinition}). In polymer quantization, the matrix element 
(\ref{DkFunctionGeneral}) can 
be expressed as

\begin{eqnarray}\label{Re-DkFunctionGeneral}
&&D_{\k}^{poly}(\Delta t) = \sum_{n}|b_{4n+3}|^2 e^{-i\Delta E_{4n+3}\Delta t},
\\ \nonumber
&=&|b_3|^2e^{-i\Delta E_3 \Delta t}\left[1+\frac{|b_7|^2}{|b_3|^2}
e^{-i(\Delta E_7-\Delta E_3) \Delta t}+ \dots\right] ~.
\end{eqnarray}
In Fock quantization, only the first term is non-vanishing. However, in polymer 
quantization there are infinitely many non-vanishing terms and from the 
asymptotic expressions (equations (\ref{SmallgC4n+3} and \ref{LargegC4n+3})) we 
can see that $b_3$ is larger than all other $b_n$ terms. Numerically we have 
shown that for the entire range of $g$, $|b_7|^2/|b_3|^2\ll1$ (Fig. 
\ref{fig:c3c7comparison}). It may also be shown that all other higher order 
coefficients are progressively smaller. Therefore, in order to simplify the 
numerical computation we restrict to the $b_3$ term only. For the purpose of 
comparison, we also plot the asymptotic expressions obtained from Eq. 
(\ref{SmallgDeltaE4n+3}-\ref{LargegC4n+3}).
\subsubsection{Coefficient $b_k$ and energy gap $\Delta E_k$}
\begin{figure}[h!]
\centering
\includegraphics[width=8.1cm,height=9.5cm]{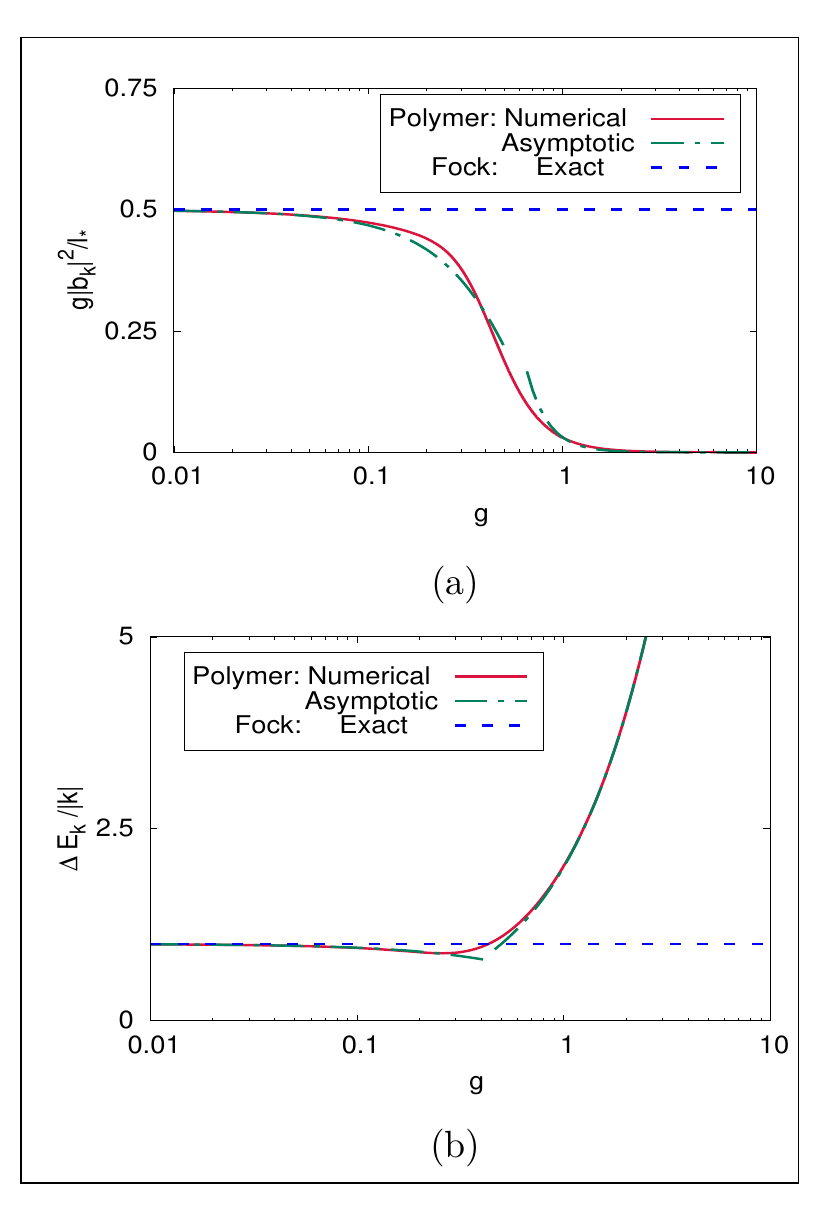}
\caption{In Fock quantization, $g |b_k|^2/\lstar = \frac{1}{2}$ and is 
represented by the dashed blue line. The solid red line represents the 
numerical results and the dot dashed green line makes use of the asymptotic 
expressions for $|b_k|^2$ in polymer quantization. 
In Fock quantization, $\Delta E_3/|\k|=1$, which is represented by the  
blue dashed line. The solid red line represents the numerical results and 
the dot dashed green line represents asymptotic expressions for polymer 
quantization. 
\label{fig:ck} }
\end{figure}

As discussed earlier, there is only one non-vanishing coefficient $b_1$ 
in Fock quantization which can be expressed in terms of dimensionless parameter 
$g$ as $|b_k|^2 \equiv |b_1|^2 = \frac{1}{2} (\lstar/g)$. 
For the  purpose of comparison, we denote the coefficient $b_3$ as 
$b_k$ and the corresponding energy gap $\Delta E_3$ as $\Delta E_k$ also for 
polymer quantization. Figure \ref{fig:ck} depicts 
$|b_k|^2$ as a function of $g$.  The energy 
gap $\Delta E_k \equiv \Delta E_3 = |\k|$ in Fock quantization and hence 
the ratio $\Delta E_k/|\k|$ is unity for all values of $g$. However, in polymer 
quantization, that ratio dips below unity and has a minima at 
$g \approx 0.26$. The behaviour of the energy gap $\Delta 
E_k$ as a function of $g$ is shown in Fig. \ref{fig:ck}.
%


\subsubsection{Two-point function}
In order to facilitate the numerical computation, we scale the two-point 
function as 
\begin{equation}
G (\Delta t,\Delta \x) = \frac{1}{4\pi^2\lstar^2} ~\tilde{G}  ~,
\end{equation}
where $\tilde{G}$ is dimensionless. Taking into account the standard regulator 
$\epsilon$, and with the help of equations (\ref{KGPropagatorDiffPM}) and 
(\ref{GPMDefinition}), the \emph{dimensionless} two-point function can be 
expressed as
\begin{equation}\label{wightmantilde}
\tilde{G}^{\epsilon} = \int_{g_{min}}^{g_{max}}dg~ 
\mathcal{T}(g,|\Delta \x|)~e^{-i p(g) - \epsilon g},
\end{equation}
where $g_{min}$ and $g_{max}$ are limits of integration which are 
used to numerically represent  $0$ and $\infty$, 
respectively. The 
above equation expresses the dimensionless two-point function on a uniform 
platform for both the Fock and polymer quantizations, respectively. 
The function $\mathcal{T}$ can be expressed in terms of 
dimensionless quantities as follows
\begin{equation}\label{ugFunction}
\mathcal{T}(g,|\Delta \x|) = 2~ \left(\frac{|b_k|^2 g}{\lstar}\right) 
~ \left(\frac{\lstar}{|\Delta \x|}\right)
~ \sin\left(g \frac{|\Delta \x|}{\lstar} \right) ~.
\end{equation}
Similarly, the 
function $p(g)$ can also be expressed in terms of the dimensionless quantities 
as
\begin{equation}\label{ugFunction}
p(g) = g~\left(\frac{\Delta E_k}{|\k|}\right)
~ \left(\frac{\Delta t}{\lstar}\right) ~.
\end{equation}
We should emphasize here that spacetime intervals $\Delta 
\x$ and $\Delta t$ are expressed in the units of $\lstar$.


In order to numerically compute the scaled two-point function $\tilde{G}$ (Eq. 
\ref{wightmantilde}) in polymer quantization, we have used $g_{min}=10^{-3}$, 
$g_{max}=10^{3}$ and the integral regulator is taken to be zero, $i.e.~, 
~\epsilon=0$. It should be noted here that a finite regulator is required in 
the Fock quantization in order to regulate the behavior of the two-point 
function. In contrast, in polymer quantization an inbuilt regulator precludes 
the need for an additional regulator. 

\begin{figure}[h!]
 \includegraphics[scale=.73]{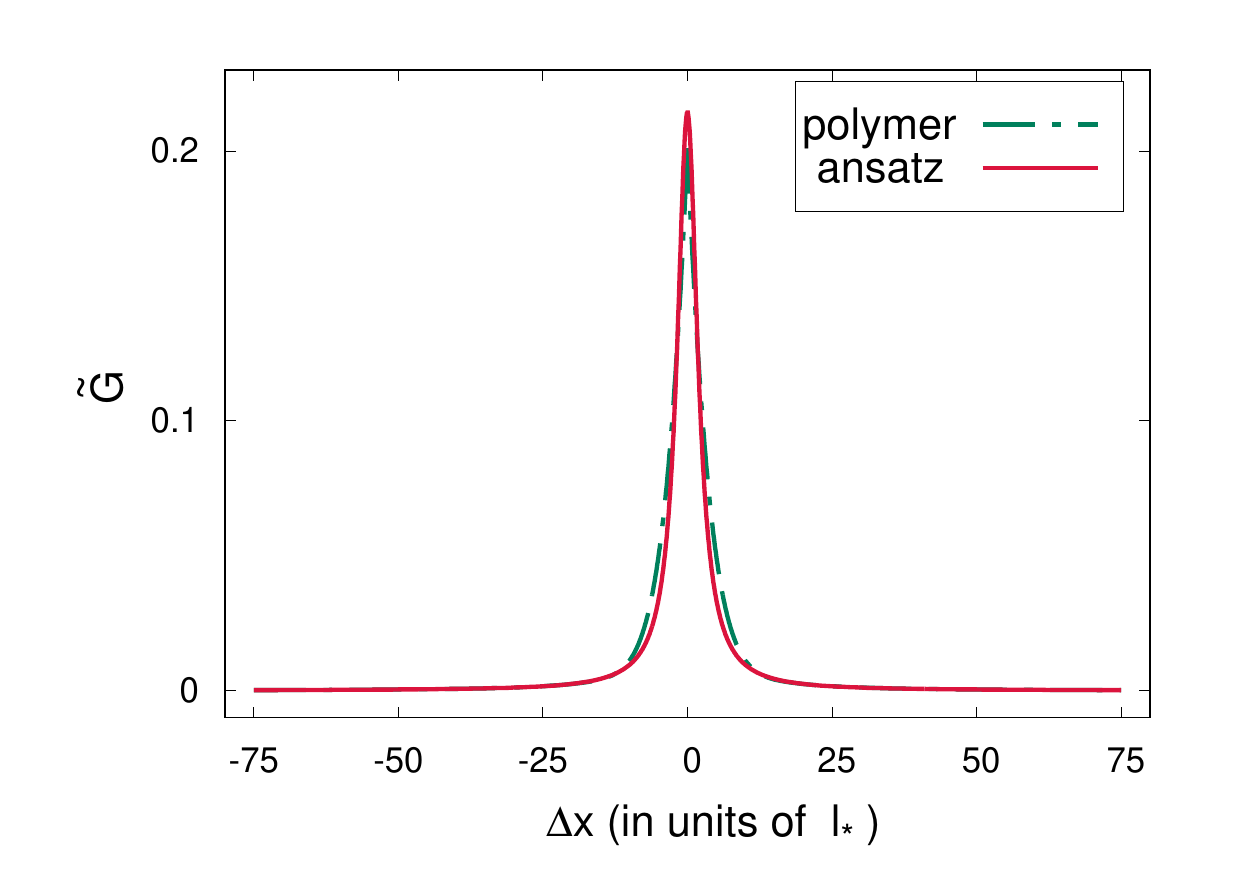}
 \caption{\label{fig:space-like-wightman} Figure depicts the scaled two-point 
function 
$\tilde{G}=4\pi^2 l_{\star}^2 G$ with respect to spatial intervals $\Delta \x$, 
when $\Delta t=0 $. The dot dashed green and solid red line represents the 
result 
obtained from polymer quantization and from the ansatz respectively. In order to 
compute the polymer two-point function, we have taken the integral regulator 
$\epsilon=0$.}
\end{figure}

\begin{figure}[h!]
\centering
 \includegraphics[width=8.5cm,height=10cm]{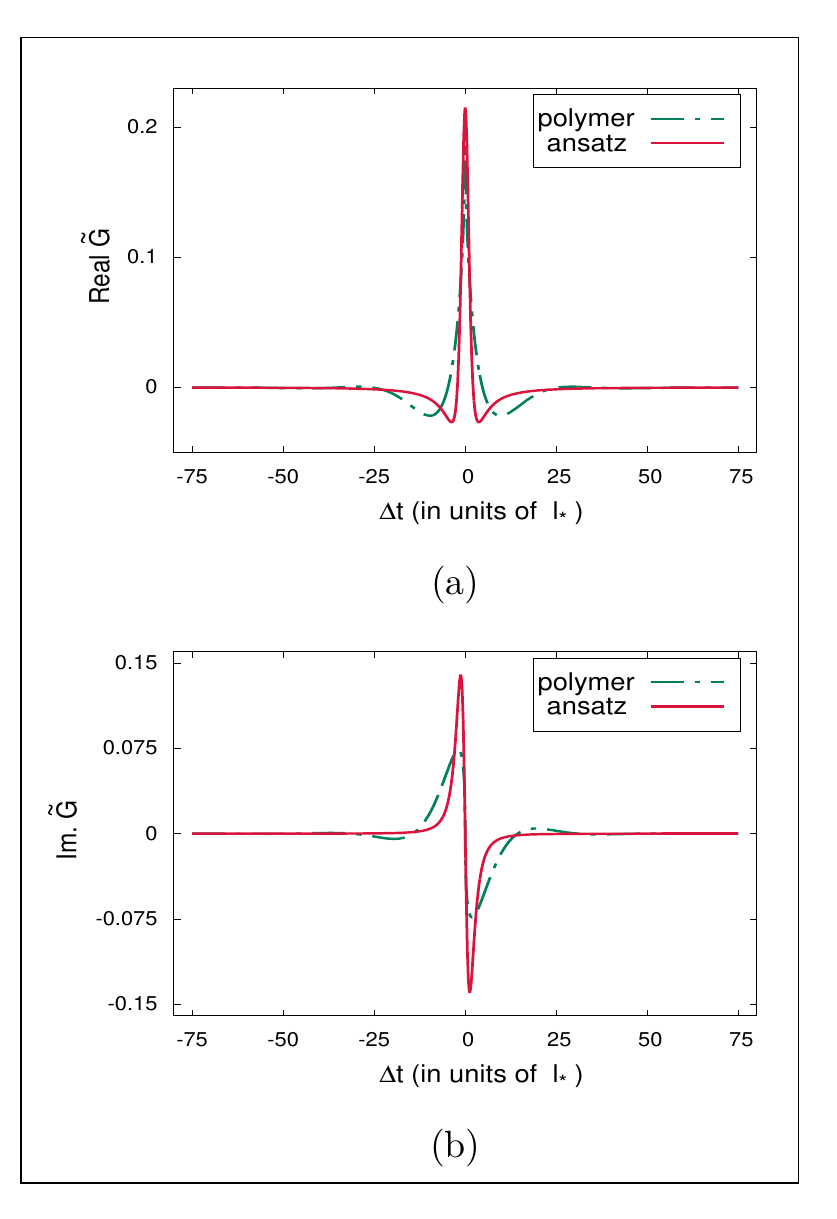}
 \caption{\label{fig:timelike-real-wightman} The sub-fig (a) represents the 
real-part and (b) represents the imaginary-part of the 
scaled two-point function $\tilde{G}=4\pi^2 l_{\star}^2 G$ with respect to 
temporal intervals $\Delta t$, when $\Delta \x\approx 0~(10^{-25})$. The dot
dashed green and solid red lines represent the result obtained from 
polymer quantization and from the ansatz, respectively. In order to compute the 
polymer two-point function, we have taken the integral regulator $\epsilon=0$.}
\end{figure}

We can see from the Fig. \ref{fig:space-like-wightman} that, for $\Delta t=0$, 
the scaled two-point function $\tilde{G}$ is real for all possible spatial 
intervals $\Delta \x$. On the other hand, for $\Delta \x=0$, $\tilde{G}$ has 
both real part (Fig. \ref{fig:timelike-real-wightman}(a)) and imaginary part 
(Fig. \ref{fig:timelike-real-wightman}(b)) for all possible temporal intervals 
$\Delta t$ . We also note here that unlike the Fock quantization, $\tilde{G}$ is 
bounded from above in polymer quantization and it converges to $\sim 0.21$ as 
both $\Delta t \to 0$ and $\Delta \x \to 0$. Analyzing these properties of  
$\tilde{G}$ and comparing with the standard form obtained from Fock 
quantization, we may conclude that in polymer quantization, there is an 
imaginary constant factor associated with $\Delta t$ which comes due to the 
standard ``$i\epsilon$'' regularization in Fock quantization. In polymer 
quantization this $\epsilon \approx 2.16$ whereas in Fock quantization, the 
limit $\epsilon \to 0$ is taken at the end of the computation. Therefore, in 
analogy with the Fock space two-point function, we can make an ansatz of the 
polymer two-point function as $\tilde{G}=1/[-(\Delta t-i~ 2.16)^2+|\Delta 
\x|^2]$.

%


\subsubsection{Unruh effect}

\begin{figure}[h!]
\centering
 \includegraphics[scale=0.73]{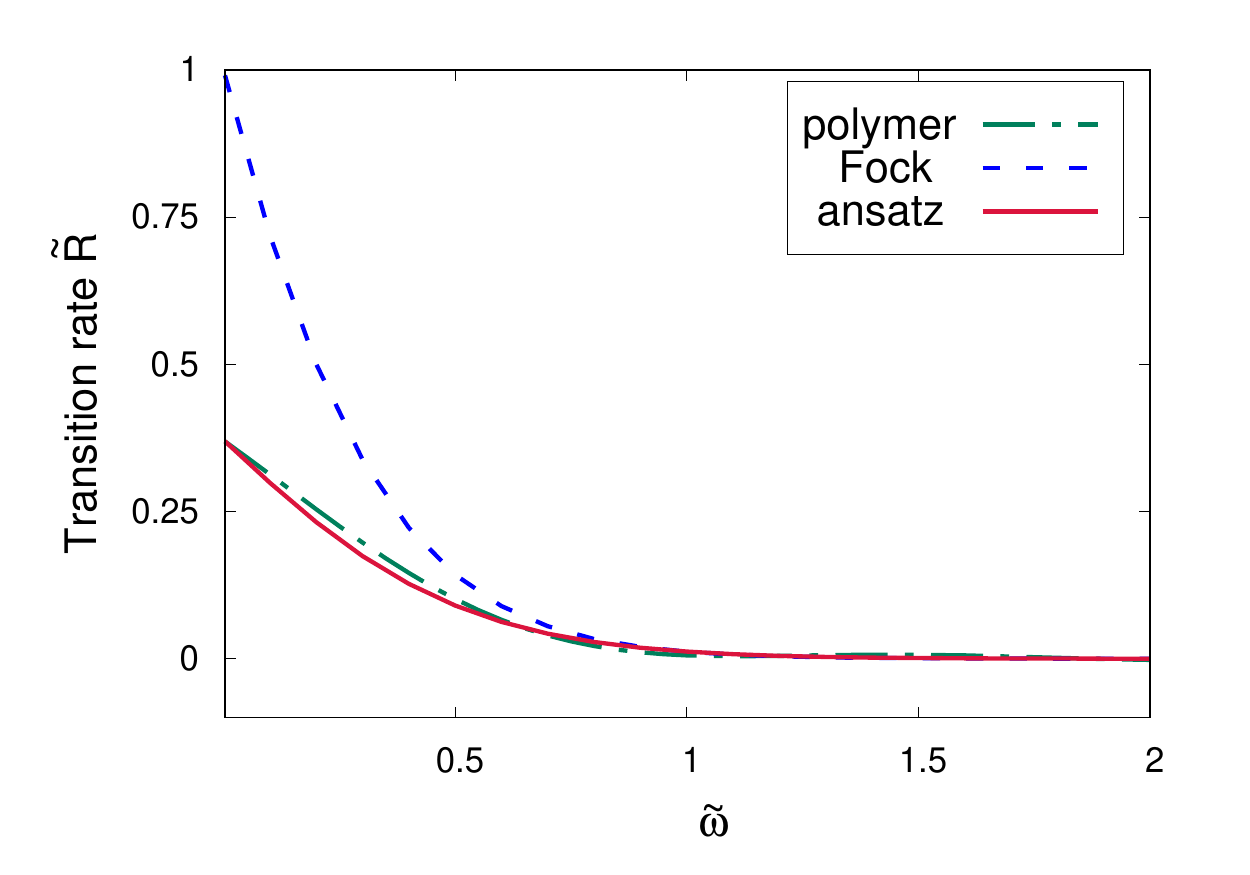}
 \caption{\label{fig:ResponseFunction}The transition rate of the Unruh-DeWitt 
detector along the Rindler trajectory, where $al_{\star}=1$. The 
dot dashed 
green, dashed blue and solid red lines represent the results obtained from 
polymer quantization, Fock quantization and ansatz, respectively. In order to 
compute the transition rate we have taken the integral regulator $\epsilon=0$ 
for polymer and $\epsilon=0.01$ for Fock quantization.}
\end{figure}

In order to numerically compute the transition rate of the Unruh-DeWitt detector 
along the Rindler trajectory $R_{\omega}(\tau,0)$ (Eq. 
\ref{DetectorTransitionRate}), we have taken $a \tau=15$ and $a=1$ in the units 
of $l_{\star}^{-1}$ and the regulator is $\epsilon=0$ for polymer and 
$\epsilon=0.01$ for Fock quantization. The Fig. \ref{fig:ResponseFunction} 
exhibits the transition rate of the detector with a scaling $\tilde{R}$ with 
respect to $\tilde{\omega}=\omega/a$, where $\tilde{R}=(a/2 
\pi)R_{\omega}(\tau,0)$.  We can see that in polymer quantization there is a 
non-thermal transition rate (dot dashed green line) which closely matches with 
the transition rate of the detector using the ansatz of the polymer-two-point 
function (red line). Therefore, we may conclude that the large value of the 
regulator $\epsilon \approx 2.16$ plays a crucial role for the non-thermal 
transition rate. 
We should emphasize here that the transition rate obtained from the polymer 
quantization has large deviations from the thermal spectrum obtained from Fock 
quantization (dashed blue line) at lower $\tilde{\omega}$ which implies 
 higher \emph{acceleration} $a$. This suggests that very high 
acceleration (comparable to Planck-length scale) would be needed to probe the 
Planck-length scale effect on the Unruh effect. It would be pertinent to note 
that the value of the regulator $\epsilon \approx 2.16$ does not depend on the 
polymer length scale. Hence, a \emph{generic cut-off} is seen to emerge in 
polymer quantization, a feature that could be probed by experiments on the 
Unruh effect \cite{Nation:2011dka,Aspachs:2010hh}.  
\section{spatially smeared detector and ``$i \epsilon$'' regularization}
\begin{figure}
\centering
 \includegraphics[width=8.5cm,height=10cm]{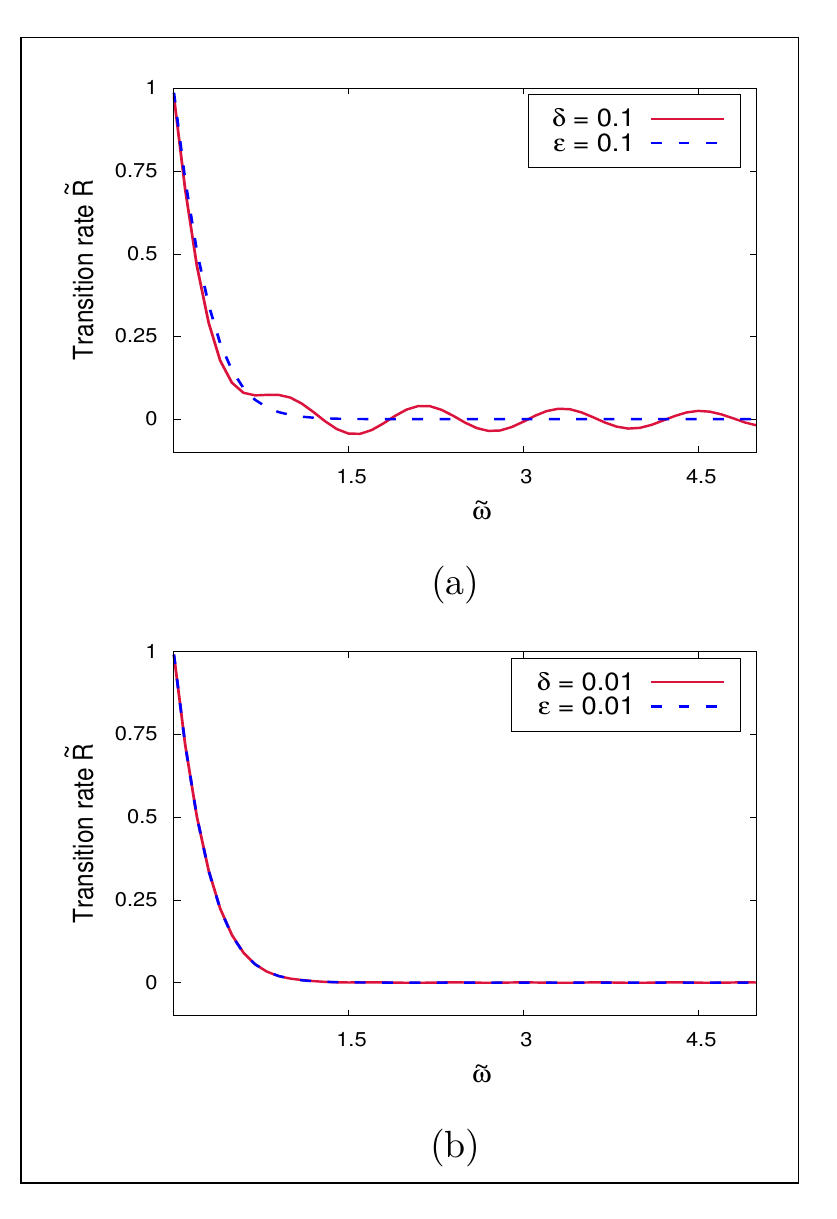}
 \caption{\label{fig:Fock-response-0-1}Transition rate of the Unruh-DeWitt 
detector along the Rindler trajectory in Fock quantization, where 
$al_{\star}=1$. The solid red line and dashed blue line represents the 
transition rate for different detector size regulator $\delta$ and the 
regulator $\epsilon$, respectively. The sub-fig (a) represents detector size  
and $\epsilon $ regulator
$\delta=0.1$, $\epsilon=0$; $\delta=0,~\epsilon=0.1 
$ and the sub-fig (b) represents detector size  
and $\epsilon $ regulator
$\delta=0.01$, $\epsilon=0$; $\delta=0,~\epsilon=0.01 $.}
\end{figure}

\begin{figure}[h!]
\centering
 \includegraphics[scale=.73]{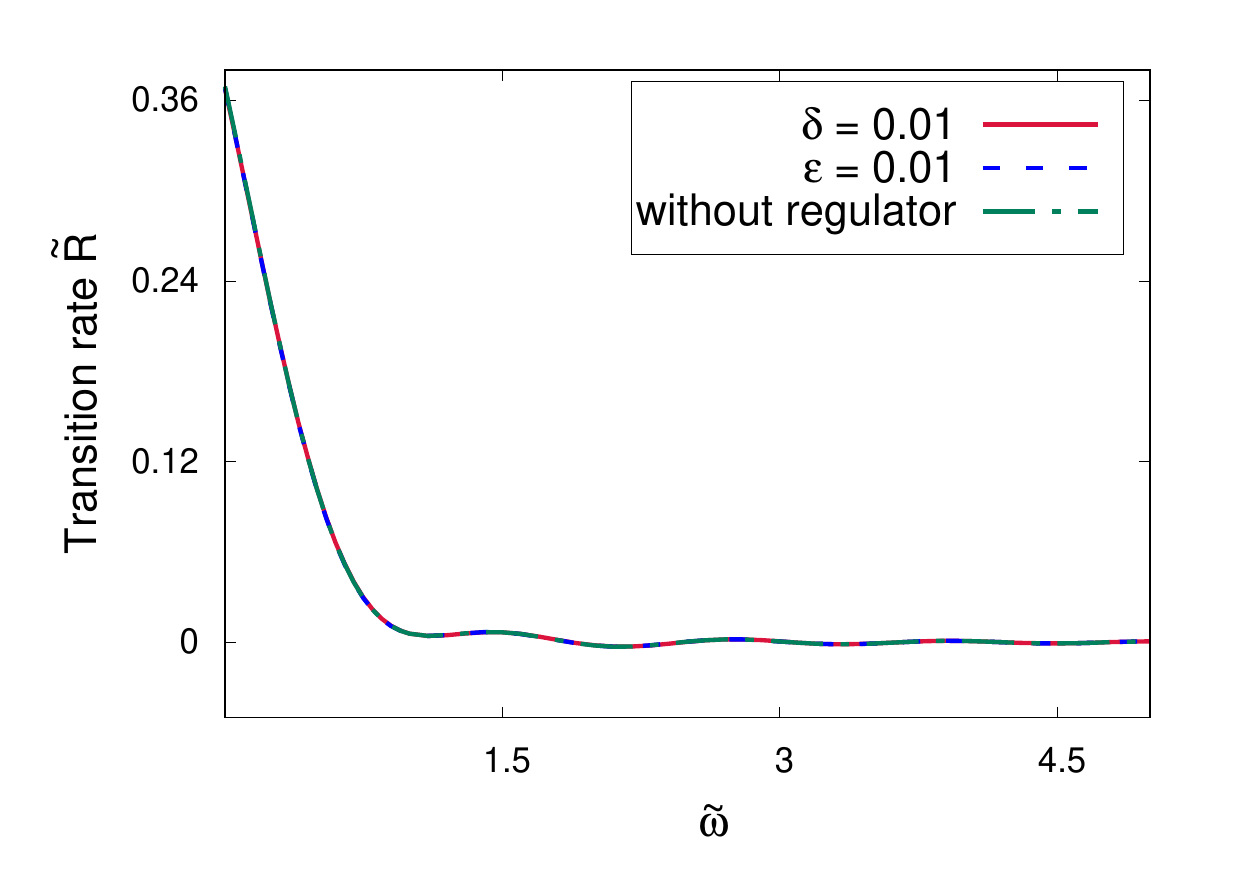}
 \caption{\label{fig:polymer-response-0-01}Transition Rate of the 
Unruh-DeWitt detector along the Rindler trajectory in polymer quantization, 
where $al_{\star}=1$. The solid red, dashed blue and dot dashed green lines 
represent the results obtained from 
polymer quantization with the detector size  and $\epsilon $ regulators 
$\delta=0.01$, $\epsilon=0$; $\delta=0$, $\epsilon=0.01$ and 
$\delta=\epsilon=0$, respectively.}
\end{figure}

Up to this point our study of the Unruh effect involves a point-like detector. 
In order to regularize the two-point function, the standard ``$i\epsilon$'' 
regularization technique is used in Fock quantization. However, there is an 
issue with Lorentz invariance and it can be taken care of by considering a 
spatially smeared detector for which the field operator will be
\begin{equation}
 \hat{\Phi}(\tau)=\int d^{3}\chi~f(\chi)\hat{\Phi}(x(\tau,\chi)),
\end{equation}
where $\tau$ and $\chi$ are the Fermi-Walker coordinates which are associated 
with the trajectory $x(\tau)$, and $f(\chi)$ is the spatial profile of the 
detector. If the spatial profile of the 
detector is 
taken as \cite{Schlicht:2003iy}
\begin{equation}
 f_{\delta}(\chi)=\frac{1}{\pi^2}\frac{\delta}{(\chi^2+\delta^2)^2}~,~\delta>0,
\end{equation}
then the two-point function in Fock quantization becomes
\begin{equation} 
G(x(\tau),x(\tau'))=\frac{1/4\pi^2}{(x(\tau)-x(\tau')-i\delta(\dot{x} 
(\tau)+\dot{x}(\tau')))^2 }~.
\end{equation}
We have numerically computed the transition rate of the detector  
considering both spatially smeared detector and point-like detector where 
``$i\epsilon$'' regularization is used. It can be seen from  
Fig. \ref{fig:Fock-response-0-1} (a) that the transition rate is much more 
insensitive to the detector size regulator $\delta$ than the standard regulator 
$\epsilon$. 
Fig. \ref{fig:polymer-response-0-01} depicts the transition rate of the 
detector, in polymer quantization. It can be seen that the 
transition rate is similar for small values of the detector size $\delta$ and 
$\epsilon$ regulators,  $\delta=0.01,\epsilon=0.0 $ and 
$\delta=0.0,\epsilon=0.01$, respectively for both quantization methods.

\section{Discussions}
To summarize, we have studied the transition rate of a uniformly accelerated 
Unruh-DeWitt detector weakly coupled to a massless scalar field for both the 
Fock as well as polymer quantizations. An essential ingredient for computing the 
transition rate of the detector is the two-point function along the detector's 
trajectory. For the case of polymer quantization, this is accomplished 
numerically. By comparing the numerically computed polymer and Fock space 
two-point functions, it is observed that the the regulator $\epsilon$ which is 
used for the standard regularization for Fock space two-point function is 
generic in the case of polymer-two-point function with a finite value $\epsilon 
\approx 2.16$. Thus, a \emph{generic cut-off} is seen to emerge in polymer 
quantization, a feature that could be probed by experiments on the Unruh effect.

Subsequently, the transition rate of the accelerated detector has been computed. 
This rate is non-thermal in polymer quantization, for high detector acceleration 
and closely matches with the transition rate using the ansatz of the polymer 
two-point function, $i.e.$ , the Fock space two-point function with the 
regulator value $\epsilon \approx 2.16$. Therefore, it follows that the large 
value of the regulator $\epsilon \approx 2.16$  leads to  deviation from the 
thermal spectrum, as obtained from the Fock quantization. The deviation 
increases as the acceleration $a$ increases. This suggests that in order to 
probe Planck scale effect on Unruh effect one needs to have a large 
acceleration. We would like to emphasize here that the value of the regulator 
$\epsilon \approx 2.16$ does not depend on the polymer length scale. 

Finally, we have also discussed the role of a spatially smeared detector on  the 
transition rate. It can be seen that the transition rate is more sensitive to 
the detector size $\delta$ than the standard regulator $\epsilon$. However, for 
small value of the $\delta$ and $\epsilon$, the transition rate is similar for 
both quantization methods.



\vspace{1 cm}

\end{document}